\documentclass[preprint,journal]{vgtc}

\vgtcinsertpkg
\ieeedoi{10.1109/TVCG.2020.3028946}

\ifpdf%                                % if we use pdflatex
  \pdfoutput=1\relax                   % create PDFs from pdfLaTeX
  \usepackage{graphicx}                % allow us to embed graphics files
  \DeclareGraphicsExtensions{.pdf,.png,.jpg,.jpeg} % for pdflatex we expect .pdf, .png, or .jpg files
\else%                                 % else we use pure latex
  \usepackage{graphicx}                % allow us to embed graphics files
  \DeclareGraphicsExtensions{.eps}     % for pure latex we expect eps files
\fi%

\graphicspath{{figures/}{pictures/}{images/}{./}} % where to search for the images

\usepackage{microtype}                 % use micro-typography (slightly more compact, better to read)
\PassOptionsToPackage{warn}{textcomp}  % to address font issues with \textrightarrow
\usepackage{textcomp}                  % use better special symbols
\usepackage{mathptmx}                  % use matching math font
\usepackage{times}                     % we use Times as the main font
\usepackage{cite}                      % needed to automatically sort the references
\usepackage{tabu}                      % only used for the table example
\usepackage{booktabs}                  % only used for the table example
\usepackage{enumerate}
\usepackage[shortlabels]{enumitem}
\usepackage{subcaption}

\usepackage{tikz}
\usepackage{textcomp}
\usepackage{hyperref}
\usepackage{lipsum}

\newcommand\copyrighttextmy{%
  \tiny\textcopyright 2020 IEEE. Personal use of this material is permitted. Permission from IEEE must be obtained for all other uses, in any current or future media, including reprinting/republishing this material for advertising or promotional purposes, creating new collective works, for resale or redistribution to servers or lists, or reuse of any copyrighted component of this work in other works.
  M. Schufrin, S. L. Reynolds, A. Kuijper and J. Kohlhammer, "A Visualization Interface to Improve the Transparency of Collected Personal Data on the Internet," in IEEE Transactions on Visualization and Computer Graphics, vol. 27, no. 2, pp. 1840-1849, Feb. 2021, doi: 10.1109/TVCG.2020.3028946.
  }
\newcommand\copyrightnotice{%
\begin{tikzpicture}[remember picture,overlay]
\node[anchor=south,yshift=10pt] at (current page.south) {\fbox{\parbox{\dimexpr\textwidth-\fboxsep-\fboxrule\relax}{\copyrighttextmy}}};
\end{tikzpicture}%
}

\onlineid{0}

\vgtccategory{Research}
\vgtcpapertype{please specify}

\title{A Visualization Interface to Improve the Transparency of Collected Personal Data on the Internet}

\author{Marija Schufrin, Steven Lamarr Reynolds, Arjan Kuijper and J{\"o}rn Kohlhammer, \textit{Member, IEEE}}

\authorfooter{
\item
 Marija Schufrin is with Fraunhofer IGD, Germany. E-mail: marija.schufrin@igd.fraunhofer.de.
\item
 Steven Lamarr Reynolds is with Fraunhofer IGD, Germany. E-mail: steven.lamarr.reynolds@igd.fraunhofer.de.
\item
 Arjan Kuijper is with Fraunhofer IGD, TU Darmstadt, Germany. E-mail: arjan.kuijper@igd.fraunhofer.de.
\item
 J{\"o}rn Kohlhammer is with Fraunhofer IGD, TU Darmstadt, Germany. E-mail: joern.kohlhammer@igd.fraunhofer.de.
}
\teaser{
  \includegraphics[width=\linewidth]{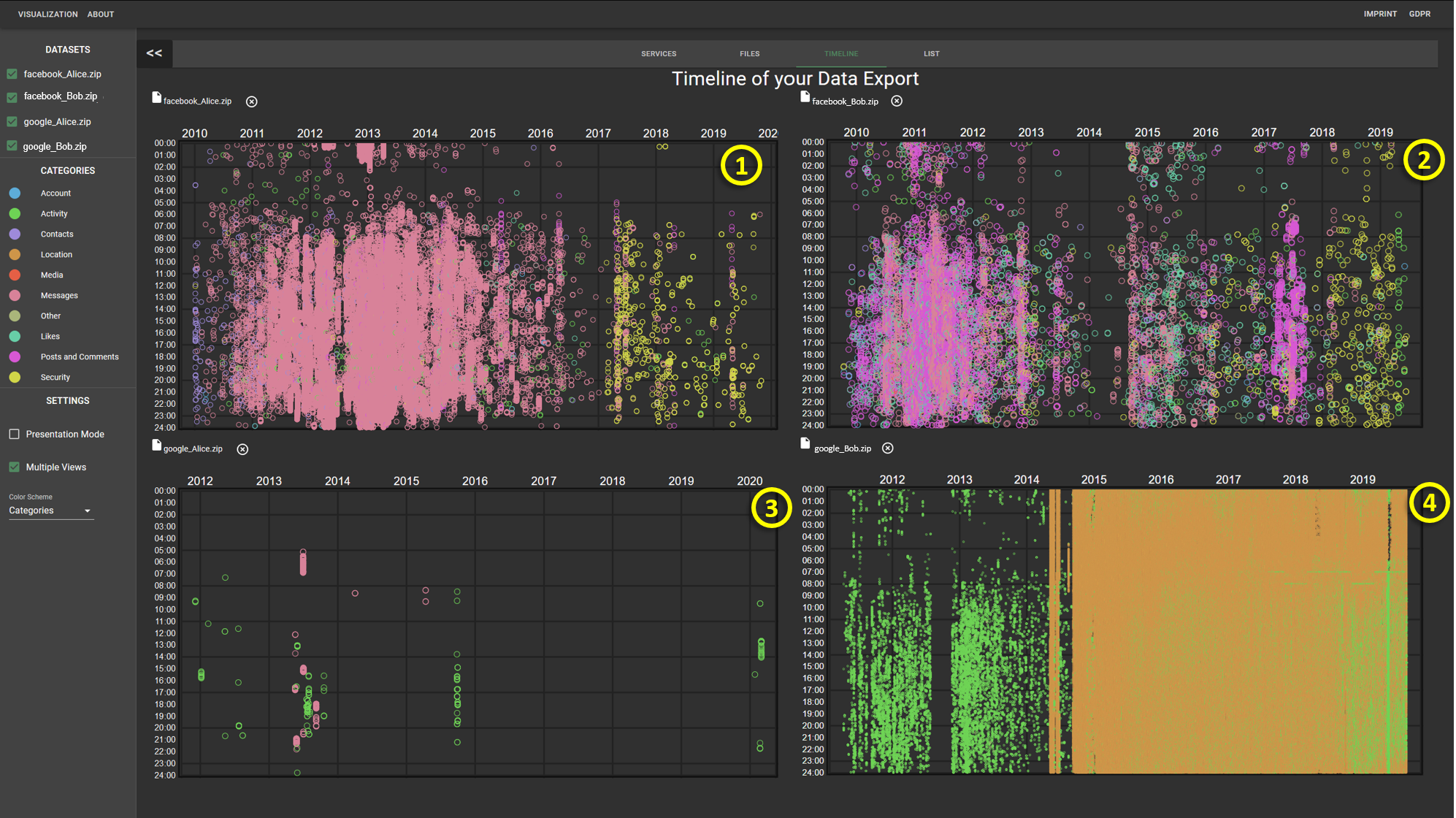}
    \caption{The \textit{TimeView} of the web interface \textit{TransparencyVis} with \textit{MultiView} mode on.
    The data elements from the GDPR data exports of two different users, each from \textit{Google} and \textit{Facebook}, are visualized in interactive scatterplots as circles over time.
    Different colors represent the categories of the data elements. Patterns can be detected and compared as described in use case 2 (see Sect. 5.2).
  }
       \label{fig:teaser}
}

\abstract{Online services are used for all kinds of activities, like news, entertainment, publishing content or connecting with others. But information technology enables new threats to privacy by means of global mass surveillance, vast databases and fast distribution networks. Current news are full of misuses and data leakages. In most cases, users are powerless in such situations and develop an attitude of neglect for their online behaviour. On the other hand, the GDPR (General Data Protection Regulation) gives users the right to request a copy of all their personal data stored by a particular service, but the received data is hard to understand or analyze by the common internet user. This paper presents \textit{TransparencyVis} - a web-based interface to support the visual and interactive exploration of data exports from different online services. With this approach, we aim at increasing the awareness of personal data stored by such online services and the effects of online behaviour. This design study provides an online accessible prototype and a best practice to unify data exports from different sources.
} % end of abstract


\CCScatlist{
  \CCScatTwelve{Human-centered computing}{Visu\-al\-iza\-tion}{Visu\-al\-iza\-tion techniques}{Privacy};
  \CCScatTwelve{Human-centered computing}{Visu\-al\-iza\-tion}{Visualization design and evaluation methods}{}
}

\nocopyrightspace

\makeatletter 
\def\@oddfoot{\hfil\sffamily\small\MakeUppercase{2020 IEEE Symposium on Visualization for Cyber Security (VizSec)}\hfil} 
\renewcommand{\ps@empty}{\renewcommand{\@oddfoot}{\hfil{\small\sffamily\copyrighttext}\hfil}} 
\makeatother

\begin{document}


\firstsection{Introduction}
%%%%%%%%%%%%%%%%%%%%%%%%%%%%%%%%%%%%%%%%%%%%%%%%%%%%%%%%%%%%%%%%

\maketitle
\copyrightnotice{}

%% \section{Introduction} %for journal use above \firstsection{..} instead

% \subsection{WHY}
% \subsubsection{Possible specials/cliffhanger}
% TODO: Use Case Location Data of corona ill persons. \\
% Nice formulation: Consider the ... bla bla corona\\

% It has been shown that individuals who belong to minorities can be identified in anonymized datasets by background knowledge attacks or homogeneity attacks~\cite{machanavajjhala2006diversity}.

% \textcolor{red}{"Background Knowledge Attack: This attack leverages an association between one or more quasi-identifier attributes with the sensitive attribute to reduce the set of possible values for the sensitive attribute. For example, Machanavajjhala, Kifer, Gehrke, and Venkitasubramaniam (2007) showed that knowing that heart attacks occur at a reduced rate in Japanese patients could be used to narrow the range of values for a sensitive attribute of a patient's disease." (Wikipedia)}

% \subsubsection{Motivation}
% digital age, changes in the last century
In the last few decades humanity has entered the digital age and became a modern information society.
It is estimated that over fifty percent of the global human population is using the Internet nowadays~\cite{itu2019statistics}.
Online services are used for all kinds of activities, like news, entertainment, publishing content or connecting with others.
But information technology enables new threats to privacy through global mass surveillance, vast databases and fast distribution networks.
By using online services, data about users is collected on a daily basis.
Companies collect data to offer more content, improve their services, gather insight about the users, or to increase the relevance of advertisements.
A few major companies offer users to connect all devices to their accounts for free.
% A few major companies offer users to connect \green{all devices to their accounts} \red{their accounts on all devices to their service} for free.
%The services offer insight for users if they track their locations, health metrics, search histories or other activities.
% Additionally, more features are added like health-related services and other activity histories.
This enables services to create an increasingly detailed profile due to the continued use of these services.
Users are often unaware of the consequences of these choices and the amount of data that is collected from them as a result.
A key point is that users lose control over the data that concerns them because they are not aware of the data that is distributed in different ways over multiple services.
% \red{One main point of this is, that users lose control over the data that is related to themselves by not being aware of data, that is spread out over different services in different ways.}
Furthermore, users cannot control exactly what happens with this data.
% \textcolor{blue}{One main point of this is, that the user looses the control over the data which could be related to the own person. 
% On the one side by not being aware of the data, what is spread out over different services in different ways and on the other side by not being able to control, what happens with this data.}
Faced with this impotence, Internet users often develop an attitude of neglect of data privacy concerns.
%%--------Möglicherweise wieder einkommentieren (MS)---------
% \textcolor{blue}{Faced with this impotence internet users often evolve an attitude of ignorance, which often results in a phenomena, which is known as the privacy paradox.
% Thereby the benefit of providing data to a service is weighted more than the possible risk of data missuses. 
% Also due to the human restricted mental capacity, people often underestimate the amount and the value of their personal data stored on the internet.}
%------------------------------------------------------------
%However, with regard to the right to privacy of each individual~\cite{udhr}, 
However, with regard to the inherent human right to privacy of each individual, 
as written in Article 12 of the Universal Declaration of Human Rights~\cite{udhr},
the ability to control the provision of one's own data to different services on the internet is of utmost importance~\cite{potzsch2008privacy}. 
\textit{Privacy} describes the right of individuals to decide how they seclude and expose information about themselves.
% It is an inherent human right as written in Article 12 of the Universal Declaration of Human Rights~\cite{udhr}.
In the context of this paper, the primary focus is on \textit{informational privacy}.
It can be described as \textit{``the right to select what personal information is known about me to what people?''}~\cite{westin1968privacy}.
%--------------------ggf wieder einkommentieren (MS)---------
% Privacy awareness should allow users to behave according to their privacy attitudes and make informed decisions.
% People that are considered privacy-aware should be conscious of their personal data.
% However, people do not behave according to their stated attitudes.
% This contradiction between stated privacy attitude and actual behavior is called the \enquote{privacy paradox}~\cite{brown2001studying, potzsch2008privacy}.
% In real situations it is hard for users to estimate the value of their personal data.
% Another important consideration is the trust a user has about a service.
% Trust is the main indicator in a users' desire and behavior to share personal data with a service~\cite{norberg2007privacy, cabinakova2016empirical}.
% Solutions are needed to enable users to make decisions based on their privacy attitudes.
%%-----------------------------------------------------------

%
% maybe value of data?
% law gdpr
% \textcolor{green}{To protect the data of users, the General Data Protection Regulation (GDPR)~\cite{EUdataregulations2016} was enacted in the European Union in 2018.
% It aims to give more control to the users by regulating how companies can collect, store and use their personal data.
% It also enables users to download and access their personal data and move it to other services.
% However the format of the exports and the huge differences from service to service make it hard for the causal internet user to get an overview over the content.}
%
% human struggle, so many services, limit on human cognition

People are using so many services today, that it is often challenging to keep track of the data they collect.
People employ different tactics to preserve their privacy.
Teenagers for example try to flood the services with random non-sensitive content~\cite{boyd2014s}.
Other try to denounce privacy threats by using common arguments such as \textit{``I have nothing to hide''}~\cite{solove2007ve}.
%%%%%%%%%%%%%%%%%%%%%%%%
% \red{
% People are using so many services today, that it is often challenging to keep track of each of them.
% This kind of personal data overload is even used as privacy preserving method. 
% Boyd~\cite{boyd2014s} found that teenagers maintain their privacy by obfuscation through flooding services with random non-sensitive content.
% A common argument to privacy threats is also \textit{``I have nothing to hide''}~\cite{solove2007ve}.
% }
%However it forces the discussion of privacy to a narrow point of view and hides the main issues of surveillance and data collection~\cite{solove2007ve}.
We argue that the main reason for such arguments and tactics is the impotence to grasp the amount and value of the personal data being collected.
Therefore, means to support the users mental access to these data collections are desirable.
We argue that visualizing such data collections in a usable way can contribute to the situational awareness of the common internet user concerning the own personal data stored at different services.
The data collection exports introduced by the GDPR, which was enacted in the European Union in 2018, proved to be a valuable resource for this aim.
The GDPR gives more control to the users by regulating how companies can collect, store and use their personal data.
It also enables users to download and access their personal data and transmit it to other services.
However, the many files and the differences of formats between and within the data exports make it difficult for casual Internet users to get an overview of the content.
% \brown{However, the multiple formats of the exports and the structural differences between services make it difficult for the casual Internet user to get an overview of the content.}
% \red{However, the format of the exports and the huge differences from service to service make it hard for the casual internet user to get an overview over the content.}
%\textcolor{green}{The data from GDPR downloads proved to be a valuable resource for this aim.}
Therefore, in this paper we present \textit{TransparencyVis}\footnote{\href{https://transparency-vis.vx.igd.fraunhofer.de/}{https://transparency-vis.vx.igd.fraunhofer.de/}}, an online accessible web tool to support a visual interactive exploration of such data exports.
%\green{\textit{TransparencyVis} is accessible online and can be tried out with one's own data.}
%An instruction, on how to access the personal data, is provided inside the tool.
%\green{Due to a front-end only implementation, the tool does not send any data anywhere.}
In a user-centered design process we identified the relevant users, data and tasks, which are also presented in this paper.
We have implemented the tool experimentally for four of the most popular online services (Google, Facebook, Instagram, Twitter).
However, the interface is extensible for other services.
Therefore we share the generalization scheme for the data exports from different services, so that the community can contribute by parsing the data exports from further services.
Our main contributions are:
\begin{enumerate} \itemsep-0.5ex
    %\item We provide a web-based prototype for a visual exploration of collections of personal data from online services aquired from gdpr data downloads. 
    \item A web-based prototype for a visual exploration of the data exports enabled by the GDPR, representing the data collections of the own personal data stored by different online services. 
    \item Characterization of the relevant users, data and tasks based on Miksch and Aigner~\cite{miksch2014matter}, as appropriate for the presented challenge to raise the situational awareness concerning personal data.
    % \item Characterization of the relevant users, data and tasks based on Miksch et Aigner \cite{miksch2014matter}, as appropriate for the presented challenge and evaluation results of the effect on situational awareness concerning personal data.
    % \item \textcolor{red}{\textbf{TODO (joern?)!!!}} Characterization of the relevant users \green{as defined by Westin (?)}, data and tasks appropriate for the presented challenge and evaluation results of the effect on situational awareness concerning personal data. \green{(or specify Design Triangle here)}
    % \red{Specification of the relevant users, data and tasks according to the data-user-task triangle. TODO: Auf designtriangle beziehen. Nicht die super charakterisierung}
    \item A unification scheme to generalize the data exports from different services, to be able to merge and compare the various data sets in one visualization.
    %\item A unification scheme to generalize the data exports from different services. Datasets from different services can be merged and compared in one visualization.
    \item Evaluation of the usability and the appropriateness of the tool after the first design iteration and lessons learned and implemented changes in the current version.
\end{enumerate}

%%%%%%%%%%%%%%%%%%%%%%%%%%%%%%%%%%%%%%%%%%%%%%%%%%%%%%%%%%%%%%%%
\section{Related Work}
%%%%%%%%%%%%%%%%%%%%%%%%%%%%%%%%%%%%%%%%%%%%%%%%%%%%%%%%%%%%%%%%

A popular research field with the goal to increase the transparency of personal data is called Transparency Enhancing Technologies (TETs)~\cite{hansen2007marrying,murmann2017tools,janic2013transparency}.
% \red{Attempts to increase either the privacy or the transparency of personal data are not new.
% Most of the approaches can be categorized either as Privacy Enhancing Technologies (PETs)~\cite{fischer2017privacy} or as the subcategory Transparency Enhancing Technologies (TETs)~\cite{hansen2007marrying,murmann2017tools,janic2013transparency}.}
% Privacy-Enhancing Technologies (PETs) are systems that protect privacy by
% \textit{eliminating or reducing personal data or by preventing unnecessary and/or undesired
% processing of personal data; all without losing the functionality of the data system}\cite{borking2001laws}.
They enable users to better understand the implications of disclosing personal data, to protect their privacy and to take an active part in the value creation of services~\cite{cabinakova2016empirical}.
TETs can be categorized into tools that enhance privacy before personal data is disclosed (ex-ante TETs) and tools that retrospectively enhance privacy once personal data has been disclosed (ex-post TETs)~\cite{fischer2016transparency}.
The approach provided in this paper can be classified as ex-post TETs.
% Einordnen unseres Ansatzes
With this approach we aim to increase the situational awareness of common Internet users with respect to their personal data, which are stored by different online services.
Thereby, the approach is to visualize the current content of the data collections that have been collected so far.
The goal is to help users reflect on their privacy attitude and their future behavior.

%Conclusion on State-of-the art
In our research of related work we have found a number of helpful approaches for visual interactive systems to increase the transparency of personal data. 
However, we have not found any approach, that addressed the visualization of the complete GDPR data exports from different services in a comprehensive view.
Some approaches use parts of the download~\cite{thudt2013visits, thudt2015visual} or are actually aiming to use a direct API of the service~\cite{fischer2016transparency}.
While there are some approaches, that provide the user with the accessibility to try the tools with their own data in their own environment~\cite{thudt2013visits, baur2010streams,fischer2016transparency}, many of the approaches either require an implementation on the server side or are not designed for personal data at all~\cite{fischer2016transparency, bier2016privacyinsight, raschke2017designing, kani2011increasing}.
We have not found any approach, where the data from multiple services could be combined and explored in one tool.
However, there are tools which support data from multiple sources~\cite{thudt2013visits, riederer2016findyou}.
While many approaches extend their data by deriving or adding further information (e.g. by machine learning, statistical information or knowledge from the outside)~\cite{do2017data, riederer2016findyou}, our focus is mainly on depicting the collection as is.
Most of the related work are appropriate for the use of a non-expert in IT.
In the following, we present the most relevant groups of related work that we have found.

%ex-ante TETs and Flow visualizations
\paragraph{Visualization of data flows}
Related approaches that also aim to visualize personal collections of data with the goal of increasing the privacy awareness are \textit{DataTrack} from Fischer-Hübner et al.~\cite{fischer2016transparency, karegar2016visualizing, angulo2015usable, fischer2013can}, \textit{PrivacyInsight} from Bier et al. ~\cite{bier2016privacyinsight}, \textit{Privacy Dashboard} from Raschke et al.~\cite{raschke2017designing} and the online interactive tool developed by Kani-Zabihi and Helmhout~\cite{kani2011increasing}.
These approaches are designed to be implemented on the server side and while they are also designed for personal data, the main focus seems to be on showing the data flows, who the data is shared with, and the details of the provided information.
%\textcolor{red}{@steven: true? (auch fischer-huebner?) The users can not try these approaches with their own data}, because they are meant to be implemented by the online services which store the data. 
%\textcolor{red}{Thorsten fragt - wozu eigentlich? Das implementieren seitens der Services...}
Our approach rather focuses on visualizing a collection of personal data to be viewed by a common internet user in an easily accessible way.
%\textcolor{brown}{Our approach mainly differs in the focus on visualizing the collection of the data rather than the flows and in the accessibility of the tool to be tried with own data by the common internet user.}

%Comparably, but focus on interfere data
\paragraph{Inferred data}
The approaches of Do~Thi~Duc~\cite{do2017data} (\textit{Dataselfie}) and Rieder et al.~\cite{riederer2016findyou} (\textit{FindYou}) also aim at visualizing personal data and thereby increasing their transparency.
Their main focus is to infer additional data using machine learning and statistical means to show what is possible to infer from the data.
%while the focus of \textit{TransparencyVis} is on visualizing the data collection as is.
Do Thi Duc uses several bar charts that show the statistical information and also uses a time line visualization
similar to the one in \textit{TransparencyVis}, but only the last seven days are visible due to their focus of collecting the data in real-time.
In \textit{TransparencyVis} the whole time span of all available data is shown.
\textit{FindYou}, on the other hand, is a location auditing tool, thus providing a more specific service.
Users can enter their own location data from three popular online services, including \textit{Instagram}, \textit{Twitter} and \textit{Foursquare}.

% Privacy Policies Vis
\paragraph{Visualizing privacy policies}
There are also approaches, that visualize privacy policies, as for example Harkous et al.~\cite{harkous2018polisis}, Tesfay et al.~\cite{tesfay2018privacyguide}, or Kelly et al.~\cite{kelley2009nutrition}.
Some mentionable but not scientific web tools for this application area are \textit{PrivacySpy}~\cite{mccain2019privacyspy}, \textit{Trackography}~\cite{ttc2016trackography}, \textit{Privacy Program}~\cite{csm2013privacyprogram}, \textit{ToS;DR}~\cite{roy2012tosdr} and \textit{useguard}~\cite{rameerez2019useguard}. 
These approaches rate privacy policies based on different assessment schemes and while they help to support users in reflecting on their privacy attitude, they differ strongly from our approach by not visualizing the actual disclosed personal data.

%Personal visualizations
\paragraph{Personal visualizations}
Some approaches visualize personal data for the purpose of
%Another group of approaches also visualizes personal data. 
%However, the actual purpose of these approaches is on
reminiscing, self-reflection and self-expression rather than for privacy awareness.
%While these are tending to motivate the user to reveal more data to the services, approaches for privacy awareness have the opposite goal.
These approaches try to gain additional value of the collected data, whereas approaches for privacy awareness try to show the value of the data collection itself.
%\red{(@Marija: Ich würd sagen nicht motivieren, aber Vorteile von Data Collection zeigen. Wohingehen privacy awareness halt den prozess transparent machen will.)}
One example for this category is \textit{Visits} from Thudt et al.~\cite{thudt2013visits, thudt2015visual}, where personal location histories are visualized in an appealing and interactive way.
Users can upload their location history from \textit{Google} and three other location based services.
%\textit{Flickr}, \textit{Moves} or \textit{Openpaths}.
Another example is \textit{LastHistory}, a work of Baur et al.~\cite{baur2010streams}, which visualizes the music listening history from the Last.fm~\cite{lastfm} service and context (photo and calendar streams) in a timeline.
Both approaches visualize an already collected data set of the users and provide the possibility to use the service with personal data, even though for very specific data collections.

%Most comparable but not scientific
\paragraph{Non-scientific tools}
%\todo{Kuerzungspotential if needed}
We found also some non-scientific online-tools, which are comparable to our approach.
For example \textit{myfbdata} developed by Do~Thi~Duc~\cite{do2017myfbdata} and the \textit{Facebook Analysis Tool} by Wolfram Alpha~\cite{wolframalpha2015facebook}.
Both were designed to visualize personal data on \textit{Facebook}, either from the data export or directly via an API. 
\textit{myfbdata} provided a map and a timeline, while the tool by Wolfram Alpha let users gain insight by providing multiple visualizations about friend circles, distributions and others.
Both tools allowed few interactions, no categorization of the data, and were designed for only one online service (\textit{Facebook}).
However, they became obsolete some years ago.
%While the tool by Wolfram Alpha was also designed only for data from \textit{Facebook}, some visualization ideas could be taken into account for further extensions of \textit{TransparencyVis}.
% Other web approaches
Beyond that, there are several other online tools, which are designed to increase the transparency of personal data on the web.
%\textcolor{brown}{Unterschied herausstellen - @Steven - *schnief* musst duuuu}
One category of these tools is the visualization of tracked user activity: e.g. \textit{re:log}~\cite{odc2013relog}, \textit{Vorratsdaten} by ZEIT Online~\cite{vorratsdaten2013zeitonline}, \textit{vds-suisse} by OpenDataCity~\cite{odc2017vds}, \textit{publicdefault}~\cite{do2018publicbydefault}, \textit{OnlineStatusMonitor}~\cite{fau2014onlinestatusmonitor}, \textit{WhatsSpy Public}~\cite{zweerink2015whatsspy}, \textit{WhatsAppAll}~\cite{kloeze2019whatsallapp}.
They visualize one or multiple static data sets to show the sensitiveness of personal data.
Other tools focus on the visualization of tracking behavior on websites, \textit{Mozilla Lightbeam}~\cite{mozilla2013lightbeam}, \textit{Netograph}~\cite{netograph2019netograph}, \textit{Trackography}~\cite{ttc2016trackography}.\\
%A different kind of tools focuses on the collection and visualization of tracking that users encounter while browsing the internet \textit{Mozilla Lightbeam}~\cite{mozilla2013lightbeam}, \textit{Netograph}~\cite{netograph2019netograph},\textit{Trackography}~\cite{ttc2016trackography}.
% von Jörn:

Thus, to the best of our knowledge there is no other approach that can provide the means to analyze personal data collections simultaneously from more than one online service in a comprehensive and transparency-enhancing way. 

%%%%%%%%%%%%%%%%%%%%%%%%%%%%%%%%%%%%%%%%%%%%%%%%%%%%%%%%%%%%%%%%
\section{Data-User-Task}
%%%%%%%%%%%%%%%%%%%%%%%%%%%%%%%%%%%%%%%%%%%%%%%%%%%%%%%%%%%%%%%%

In this section we present the targeted data, user and tasks according to Miksch and Aigner's design triangle~\cite{miksch2014matter}.

\subsection{Data} \label{sec:data}
\subsubsection{GDPR downloads}
In the scope of this research topic we are focusing on personal data that is collected on the Internet.
Personal data is \textit{``any information related to an identified or identifiable natural person''}~\cite{EUdataregulations2016}.
It is primarily provided by users to online services simply by using them.
In recent years, the Internet rights for users were strengthened by the introduction of the Californian CCPA or the European GDPR~\cite{EUdataregulations2016}.
The latter provides European citizens with Article 15, the right of access, i.e. they can request a copy of their personal data, a \textit{data export}.
It also includes Article 20, the right to data portability, with which they can use their data export for their own purposes across other services.
It also requires the service to deliver the data export in a structured, commonly used and machine-readable format.

During our research, we investigated the data export request on several services and found large differences among the retrieval process.
% Automated/Email
While most services employ an automated data export, some require users to contact the support via email and identify themselves with an image of their passport.
% Duration
Further, the retrieval process varies in the duration of the time till the export is created.
%GDPR states that services should respond without undue delay and within one month.
For some services the duration depends on the size of the data export or the current workload of that service, however, a few services need several days to weeks to generate the data export.
% File Types
Most data exports we encountered were available as a zip archive and contained many different file formats, including json, js, csv, html, tcx, vcf, ics and others.
As each file contained data about various topics, they all had an individual data structure and only occasionally used reoccurring data types.
Some services used special encoding such as UTF-8 encoded strings, JavaScript files with an exported variable that contains the JSON data, or included data which purpose or context could not be identified.
These files and data structures were almost never documented by the service, only the Twitter data export provided a documentation.
It should be noted that some services allow choosing between multiple file formats, in most cases JSON and a HTML variant that allows for easier viewing.
% compelteness
Some data that was available on the website of the service was not included in the data export, but it was mostly miscellaneous data or newer features which were not added yet.
% Table
%\textcolor{brown}{A table with a list of different most popular services investigated according to these characteristics can be found in \autoref{fig:online_services} or
%the accompanying documents.}\\
% How did we select them
We selected services which are popular among users, have an automated and simple data export request feature, have a short duration to generate the data export, and allow easy maintenance.
% Which ones
We therefore decided for \textit{Facebook}~\cite{fb2019fb}, \textit{Google}~\cite{google2019google}, \textit{Twitter}~\cite{twitter2019twitter} and \textit{Instagram}~\cite{insta2019insta} in our initial prototype.

\subsubsection{Generalization}
\begin{figure} [tb]
      \centering
      \includegraphics[clip, trim=11.3cm 1.45cm 11.9cm 1.1cm, width=0.9\columnwidth]{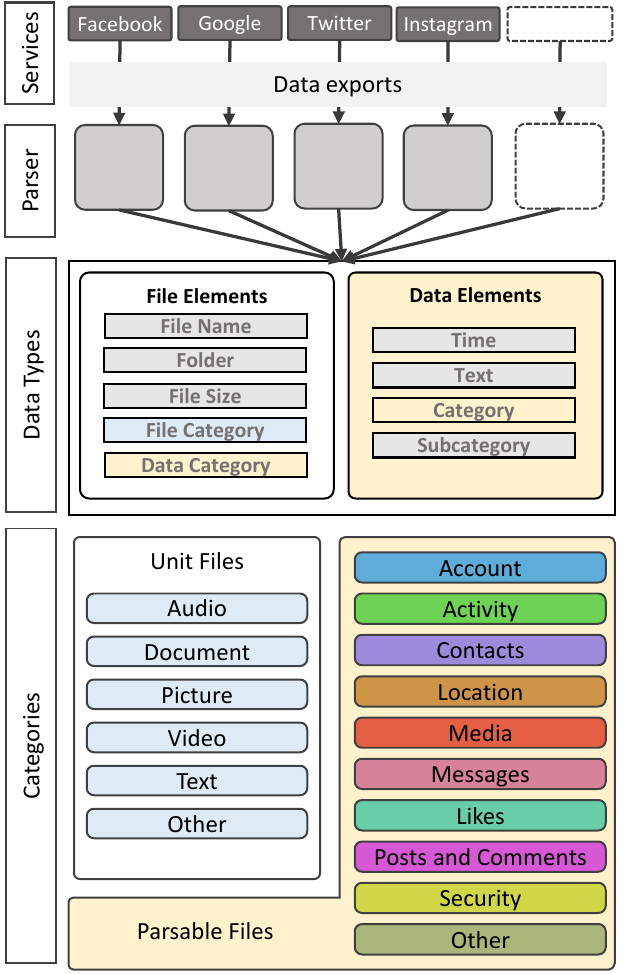}
      \caption{Unification scheme: Data exports from different online services are unified to a defined scheme by a specific set of parsers. For each service an own parser must be defined. The unification results in two data types: \textit{file elements} and \textit{data elements} as well as an assignment of the elements to a category from the defined set of categories.
      }
      \label{fig:scheme}
  \end{figure}

% intro
The data export comprises several folders that contain the data of certain parts or features of the service.
In those folders are sub folders and files in multiple file formats.
% data exchange
Some files are in a common data format, like json, while others can contain images, videos, documents or binary data which might not be known before.
%Some file formats are \textcolor{brown}{machine-readable (Steven?)} and contain data that can be parsed, while others can be binary data that might not be known a-prior.
% \textcolor{brown}{To be able to show all elements in the data export, we used the files in the data exports themselves as a data format.}
Due to the high variation of the content and the structure of the data exports, we defined a unification scheme with the goal to simplify the data and to make it comparable.
The overall unification scheme is shown in \autoref{fig:scheme}.
%\textcolor{red}{Alternativ von Thorsten:To generate an overview over data from so many different format, we categorize... }
Based on our observations of the data formats we defined two types of data for our visualizations: \\

\textbf{File elements}:  A file element represents files, which are contained in the data export. This can for example be a video, image, other archive or a machine-readable document. The files are categorized based on their file extension to make it easier for users to understand the files' purpose. The main attributes of this type of data are:
\begin{itemize}\itemsep-0.5ex
    \item \textbf{File Name} - messages.json
    \item \textbf{File Category} - Picture, Video, Audio, Text, Document, Other
    \item \textbf{Folder} - messages/
    \item \textbf{File Size} - 5 MB
    \item \textbf{Data Category} -  Messages, Security ...
    %\item \textbf{Data} -  parsed Data Elements
\end{itemize}
%(Siehe Kommentar beim nächsten itemize)

\textbf{Data element}: 
%data element represents a part of the data that could be identified in the json files contained in the ZIP folder 
Data elements represents chunks of data, which could be identified in the machine-readable files contained in the data export. 
Most of the machine-readable data was given in a list or array with individual elements that contain multiple relevant attributes.
Most elements are certain events that happen within the service.
For example, account creation, password changes, sending messages, accepting friend requests, visiting an URL, using search, liking a page and others.
After documenting several machine-readable files from multiple providers, we created the following attributes for this data type:
%The main attributes of this type of data are:
\begin{itemize}\itemsep-0.5ex
    \item \textbf{Time} - 2019-01-01 12:34:56
    \item \textbf{Text} - Person says: ``Hello World''
    \item \textbf{Category} - Messages, Security ...
    \item \textbf{Subcategory} - Chat with Person B, Chat with Person C, ...
\end{itemize}

Finally - in order to support pattern exploration and a comparison between data sets from different online services or users, a set of ten categories has been derived so that each data element could be classified according to these categories (see also \autoref{fig:scheme}):
\textit{Account} (any data related to the users' account), \textit{Activity} (any data that is collected passively from users), \textit{Contacts} (any data that contains contact addresses or friends lists or similar), \textit{Location} (any location-oriented data), \textit{Media} (any data that primarily describes media data from the user), \textit{Messages} (any communication data), \textit{Posts and Comments} (any posts or comments from the user), \textit{Security} (any security related data such as logins or IP addresses), \textit{Other} (any data that does not fit the other categories). 
File elements which contain data elements have the same value for the attribute data category as the contained data elements.
These categories can universally be applied for different services, so that a combination and/or comparison of data from different services is eased as well.

% %%%%%%%%%%%%%%%%%%%%%%%%%%%%%%%%%%%%%%%%%%%%%%%%%%%%%
\subsection{Users: The ordinary Internet user}
%\todo{Jörn fragen!}
The proposed approach has primarily been designed with the ordinary Internet user in mind.
% \brown{The main goal is to enable a usable insight into the personal data collections about a user stored by different online services.(Sagt nix über den Nutzer aus)}
% \textcolor{blue}{To satisfy the user description in the design triangle, we classify this user group according to two attributes: privacy concern and technical skills.}
For the user description in the design triangle, we characterize the user group by looking at two attributes: privacy concern and internet skills.
With respect to the privacy attitude, Westin~\cite{westin1991harris} defines three main groups of users based on their privacy concern index.
%Westin~\cite{westin1991harris} categorized the users privacy attitudes based on their privacy concern index into:
%\red{
 \begin{itemize}\itemsep-0.5ex
 \item \textbf{Fundamentalist}: A person that is distrustful of data collection by organizations and cares about privacy.
 %cares a lot about privacy and data collection %\textcolor{brown}{@steven von Jörn: and is very careful with using online services [???]}.
 \item \textbf{Pragmatist}: A person that weighs the benefits against the intrusiveness of data collection and believes that organizations should earn their trust rather then automatically have it.
 %that knows and cares about privacy and data collection.
 %\textcolor{brown}{@steven: , but ... [Da scheint noch was zu fehlen].but has a rather open attitude regarding online tools and services.}
 \item \textbf{Unconcerned}: A person that is trustful of organizations collecting personal data.
 % that does not care about privacy or data collection.
 %\item \textbf{Unmotivated User}: A person that falls under the privacy paradox and needs incentive to change.
 \label{list:concepts:users}
\end{itemize}
% (@Marija: Ich habe die Beschreibungen umgeschrieben anhand dem einer Westin Metaanalyse, passt das so?)
%}
We see benefits from the ability to gain visual insight into their own data stored by different online services for each of these groups.
%Furthermore, we assume all user groups to have \green{average digital skills~[TODO: REF!]} \red{average technical skills}.
Furthermore, we assume that all user groups have sufficient digital skills \cite{van2014measuring} to use online services such as Google, Facebook, Instagram and Twitter.
The evaluation results show that the usability is appropriate for the evaluated user groups. However, an evaluation of privacy and visualization skills against the effectiveness of these tools would be a valuable future work.

% %%%%%%%%%%%%%%%%%%%%%%%%%%%%%%%%%%%%%%%%%%%%%%%%%%%%%
\subsection{Tasks} \label{sec:tasks}
The main goal of our visualizations is to provide a comprehensive insight into the collection of personal data stored by different online services. 
This collection is represented by the data export, that can be requested from the services as guaranteed by the GDPR.
With this we aim to support the situational awareness of one's personal data on the Internet.
According to Endsley~\cite{endsley2000theoretical} situational awareness consists of three stages: \textit{perception}, \textit{comprehension}, \textit{projection}.
% Comparably Pötzsch et al. ~\cite{potzsch2008privacy} describes privacy awareness as encompassing \textit{attention}, \textit{perception}, and \textit{cognition}.
%Comparably Pötzsch et al. ~\cite{potzsch2008privacy} apply the theory of situational awareness on the context of privacy.
Applied to the context and data considered in this paper, the following three main goals can be defined.
%At this state of development we have focused on the first two stages \textit{perception} and \textit{comprehension}.
\begin{enumerate}[start=1,label={\bfseries \arabic*:}] \itemsep-0.5ex
    \item \textbf{Support Perception:} Support the investigation of the distribution of own data elements with regard to information type, time and the service by which it is stored.
    \item \textbf{Support Comprehension:} Support the identification of possibly sensitive information
    \item \textbf{Support Projection:} Increase the attention for the users current and future online behavior.
\end{enumerate}

Based on these goals we have identified the following tasks, in that our approach should support:

 \begin{enumerate}[start=1,label={\bfseries T\arabic*:}] \itemsep-0.5ex
     \item OVERVIEW of all data elements contained in the exported data collection (\textit{perception})
     \item INSPECT the details of each data element (\textit{comprehension})
     \item RELATE the data elements to services (\textit{perception and comprehension})
     \item RELATE the data elements to time (\textit{perception and comprehension})
     \item COMPARE data between services and time periods (\textit{perception and comprehension})
     \item EXPLORE possible patterns and information resulting from aggregation of the data (\textit{comprehension})
     \item REFLECT on the personal value and perceived sensitivity of the revealed information (\textit{projection})
    %  \item (possibly adding) RELATE the data elements to kind of data (\textit{perception and comprehension})
\end{enumerate}

Through the overview of the whole data collection, the users should gain a first insight into the data. 
At this stage the users might have already identify unexpected data elements.
Users can inspect the details of the data element to determine how confidential or critical the information really is to them.
By relating the data elements to the context of time or exploring different services, the users should gain an additional perspective on the value of the provided data.
Furthermore, patterns and unexpected information resulting from bringing together different data can be identified.
Finally, the active reflection of users on the personally perceived sensitivity of the data should increase the awareness for the value of the stored data.
While we defined the tasks mainly based on the three defined goals, we argue that the tasks are beneficial for all three user sub-groups. However, there might be different effects on the different sub-groups. For example, while the \textit{Fundamentalist} might use \textbf{T5} to detect sensitive information resulting from aggregation, the \textit{Unconcerned} might use \textbf{T5} to reminisce or self-reflect. On the other hand, the latter might lead to a higher awareness of their own data as a side effect.

\subsection{Design Requirements}\label{sec:requiremtens}
For the visualization solution itself the following requirements (\textbf{R1}-\textbf{R7}) have been derived based on the above data, user and task identification.
To increase the willingness of the user to use our tool, we additionally added three system related requirements \textbf{R8}-\textbf{R10}. These requirements are in line with the requirements for privacy awareness supporting tools proposed by Pötzsch et al.~\cite{potzsch2008privacy}. With these we mainly aimed to ensure that the evaluated effect on the users experience results from the real inspection of the own data and not from a mockup, which we believe, makes a huge difference.
% \textcolor{red}{Additionally, we set two general requirements on the system which aims to support privacy awareness as proposed by Pötzsch et al.~\cite{potzsch2008privacy}.} Nevertheless the proposed tool is a prototype, we see the requirements (\textbf{R8}-\textbf{R9}) as crucial factors to with regard to the willingness to use the tool and the effect on the privacy awareness of the users.

%%%%%................%%%%%
% Additionally, we set some general requirements on the system which aims to support privacy awareness as proposed by Pötzsch et al.~\cite{potzsch2008privacy}. While the proposed tool is a prototype, we think, that these four requirements have a huge influence on whether the user is willing to use the interface at all (\textbf{A1}-\textbf{A4}).
%%%%..................%%%%
\begin{enumerate} [start=1,label={\bfseries R\arabic*:}] \itemsep-0.5ex
   \item A view which shows all elements contained in the export at once (\textbf{T1})
   \item Zoom and filter, details for each data element on demand (\textbf{T2})
   \item Ability to upload data from different online services (\textbf{T3})
   \item Timeline layout for data with a time attribute (\textbf{T4, T5, T6})
   \item Visual categorisation by type of data to support the pattern exploration process  (\textbf{T5, T6})
   \item Display multiple data sets at the same time (\textbf{T5})
   \item Functionality to evaluate the perceived sensitivity of a piece of information (\textbf{T7})\\
   \item Own data, not just demo data
   \item No invasion to privacy by the prototype itself
   \item Understandable for non-experts in IT and visualization 
  % \item \textcolor{red}{Maintain interactive performance} \todo{Steven/Jörn? klären}
\end{enumerate}

%%%%%%%%%%%%%%%%%%%%%%%%%%%%%%%%%%%%%%%%%%%%%%%%%%%%%%%%%%%%%%%%
%\section{Visualization and Design Rationale}
\section{TransparencyVis} \label{sec:transparencyvis}
%%%%%%%%%%%%%%%%%%%%%%%%%%%%%%%%%%%%%%%%%%%%%%%%%%%%%%%%%%%%%%%%
% \textcolor{red}{TODO: Design Rationale zu jeder Visualisierung! (bsp von Jörn: "We decided for a scatterplot to allow a display of each single data element, while being able to perceive general trends."}

% Chapter 4.3.2
In this section, we present our prototype \textit{TransparencyVis}.
First we will explain the infrastructure and the main technologies we used in our prototype.
Then, we describe the visualization components and demonstrate how \textit{TransparencyVis} can be used in practice along some use cases.

\subsection{Infrastructure and Technology}
% A1-A4
\textit{TransparencyVis} is implemented as a web application that primarily runs on the client side.
The interface is written in \textit{TypeScript} and \textit{React.js}, and for the visualizations we use the JavaScript library \textit{d3.js}.
%It is written in \textit{TypeScript} and the user interface is created with \textit{React.js}, \textit{Redux} and \textit{Material-UI}. For the visualizations we use the JavaScript library \textit{d3.js}. 
These technologies enable the implementation of an interface with interaction paradigms familiar to the common Internet user (\textbf{R10}). 
To meet \textbf{R8} and to enable the users to explore the tool with their own data, we have implemented an upload and parsing mechanism for four exemplary, but well-known, services. %\brown{(\textit{Google}, \textit{Facebook}, \textit{Twitter}, \textit{Instagram}).
%The mechanism can, however, be extended to other services. (Wurde alles schonmal genannt?)
To ensure \textbf{R9} we decided to avoid any unnecessary connections to the server. 
Therefore, instead of uploading the data to a server, the processing is done in a web-worker thread in the browser to fulfill the privacy aspect while still being interactive. 
%as stated by \textcolor{red}{\textbf{R11}}.
%\brown{To reduce the complexity for the user as far as possible, \textit{TransparencyVis} automatically detects the service of the data set. (Die Info steht im nächsten Satz schon)}
When a data export is selected, the contents are extracted and the service is automatically detected, to reduce the complexity for the user as far as possible.
%When a data export is selected, the contents are extracted and analyzed to find the matching service.
As defined in \autoref{sec:data}, each service has its own parser for each parsable file that is used to extract the relevant data from a JSON, or other, file to the data elements.
The structure of JSON files is documented by \textit{TypeScript typings} to facilitate the extension and maintenance of the application.
Additional services can be added by implementing a parser for their data export structure. 
%\red{These parsers indicate how the data can be converted from a JSON, or other, file to the data element structure}.
%\textcolor{red}{Thorsten schlägt vor, Ruleset Beispiel zu zeigen, macht das Sinn?}

\subsection{Visualizations and Interactions}

Based on the requirements stated in \autoref{sec:requiremtens} we developed a collection of views (see \autoref{fig:screenviews}) to support the users and their tasks.
The two main views are the \textit{FileView} (b) and the \textit{TimeView} (c). They are complemented by the  \textit{Data Page} (a) and the \textit{ListView} (d).
The \textit{FileView} is mainly based on a TreeMap~\cite{shneiderman1998tree} and is primarily meant to enable the user to get an overview of all file elements contained in the export at one glance (\textbf{R1}).
The \textit{TimeView} is mainly designed as a scatterplot~\cite{cleveland1984many} with the temporal aspect of the data elements.
%While time is one-dimensional, the repetitive cycles are considered and split into two dimensions.
%two time axis for "day" and "hour of day".
%carr1987scatterplot
It contains time-dependent data and is primarily meant to explore patterns and time relations (\textbf{R4}). 
The \textit{ListView} displays all data elements in a list.
Additionally, users can rate the perceived sensitivity for each data element to support reflection (\textbf{R7}).
The common process is as follows:
The users start by retrieving their personal data from the online services and dragging the zip archive into the \textit{Data Page}.
%We provide instructions to download the data from the four services on the \textit{Data Page} of \textit{TransparencyVis}.
% The users insert the received zip archive into the tool by dragging and dropping it to the \textit{Data Page}.
Multiple zip archives from different services can be inserted at once (\textbf{R3}, \textbf{R6}).
The user proceeds by going to the \textit{FileView}, where the user can explore the files contained in the data export.
Further they can explore the temporal data in the \textit{TimeView} and finally have a look at the details in the \textit{ListView}.
However, the user can also switch between the views as desired.
The sidebar contains the ten categories, as described in \autoref{sec:data} with the mapped color (\textbf{R5}) which is the same for all views.
The data in each visualization is mapped to the color of their assigned category.
The \textit{FileView} has an additional category \textit{Files}.
The legend list in the sidebar can also be used to filter each category (\textbf{R2}). 

\subsubsection{Data Page}

\begin{figure*}[ht!]
    \centering
    \begin{subfigure}[b]{0.49\textwidth}
        \includegraphics[width=\textwidth]{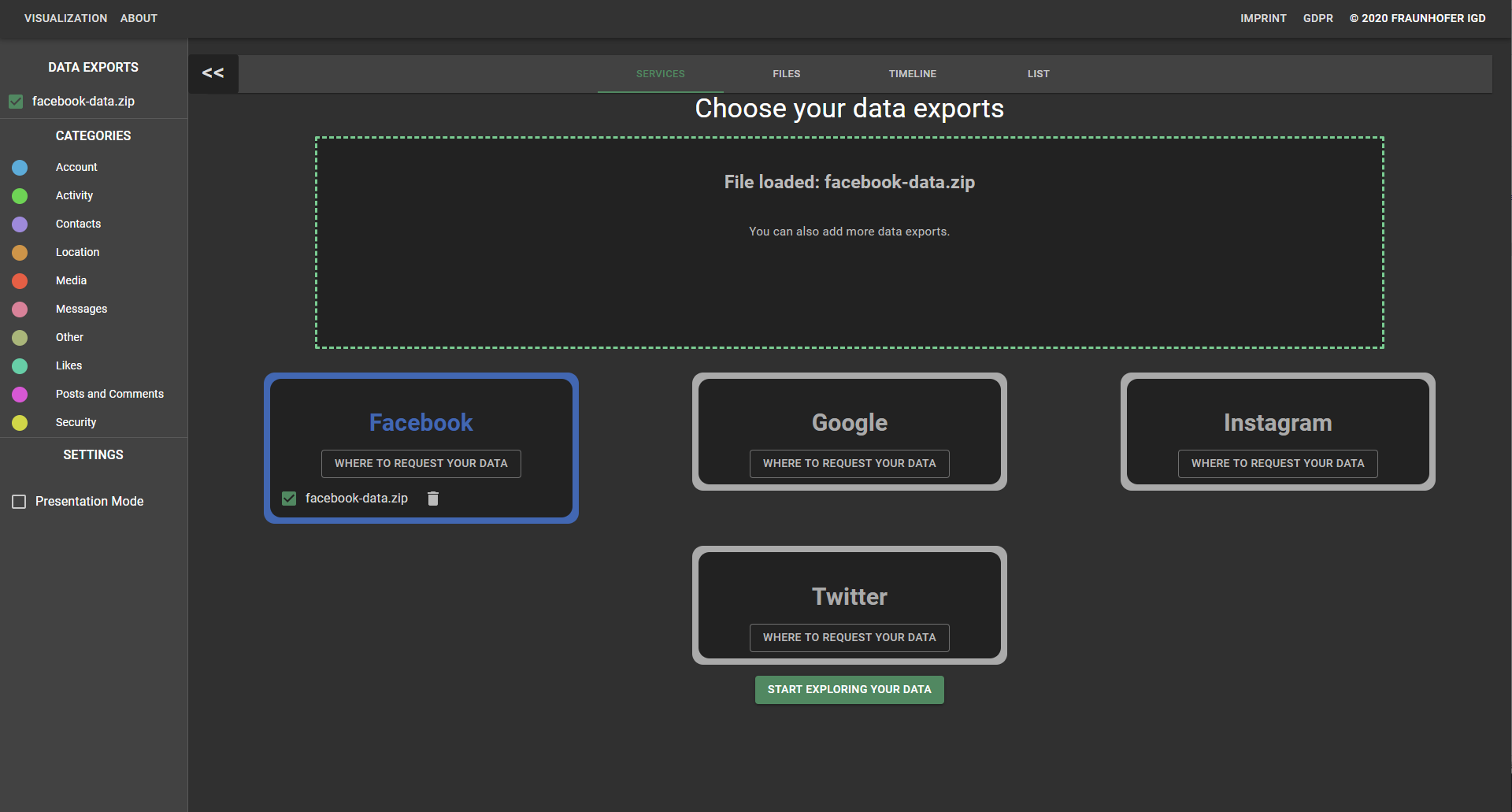}
        \caption{Data Page}
     \label{fig:screenviews:data}
    \end{subfigure}
    \begin{subfigure}[b]{0.49\textwidth}
        \includegraphics[width=\textwidth]{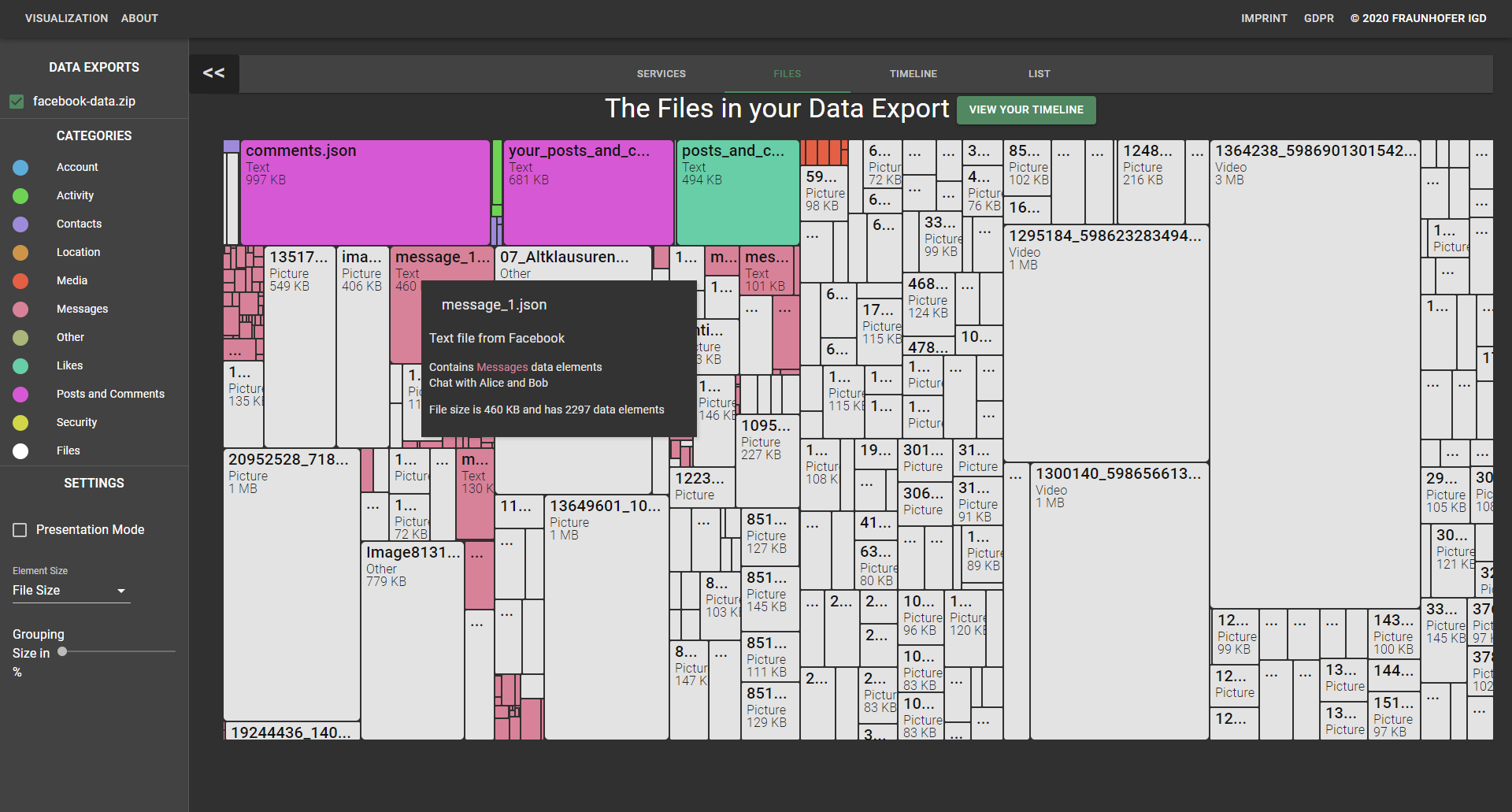}
        \caption{FileView}
     \label{fig:screenviews:file}
    \end{subfigure}
    \begin{subfigure}[b]{0.49\textwidth}
        \includegraphics[width=\textwidth]{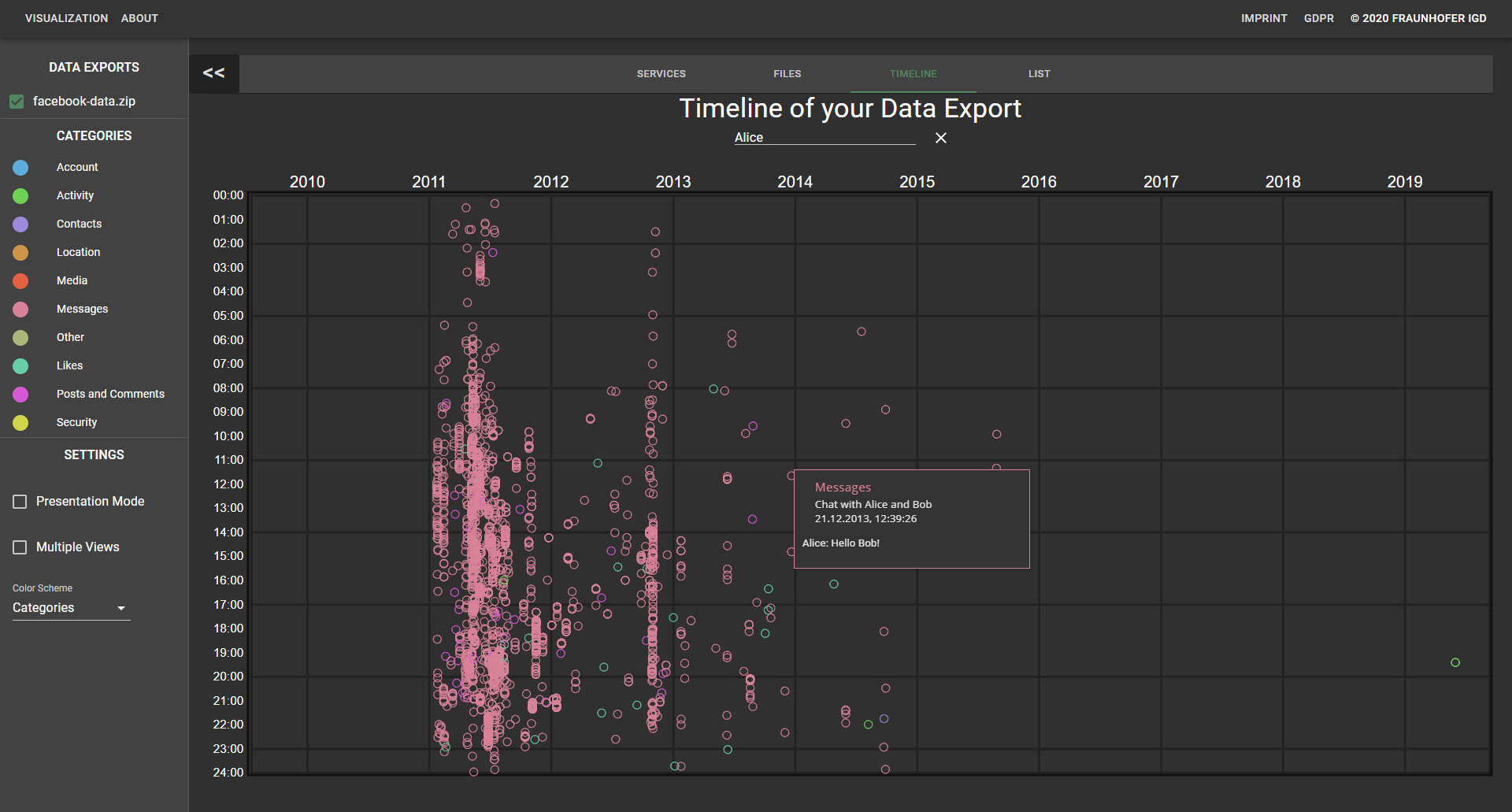}
        \caption{TimeView}
     \label{fig:screenviews:time}
    \end{subfigure}
    \begin{subfigure}[b]{0.49\textwidth}
        \includegraphics[width=\textwidth]{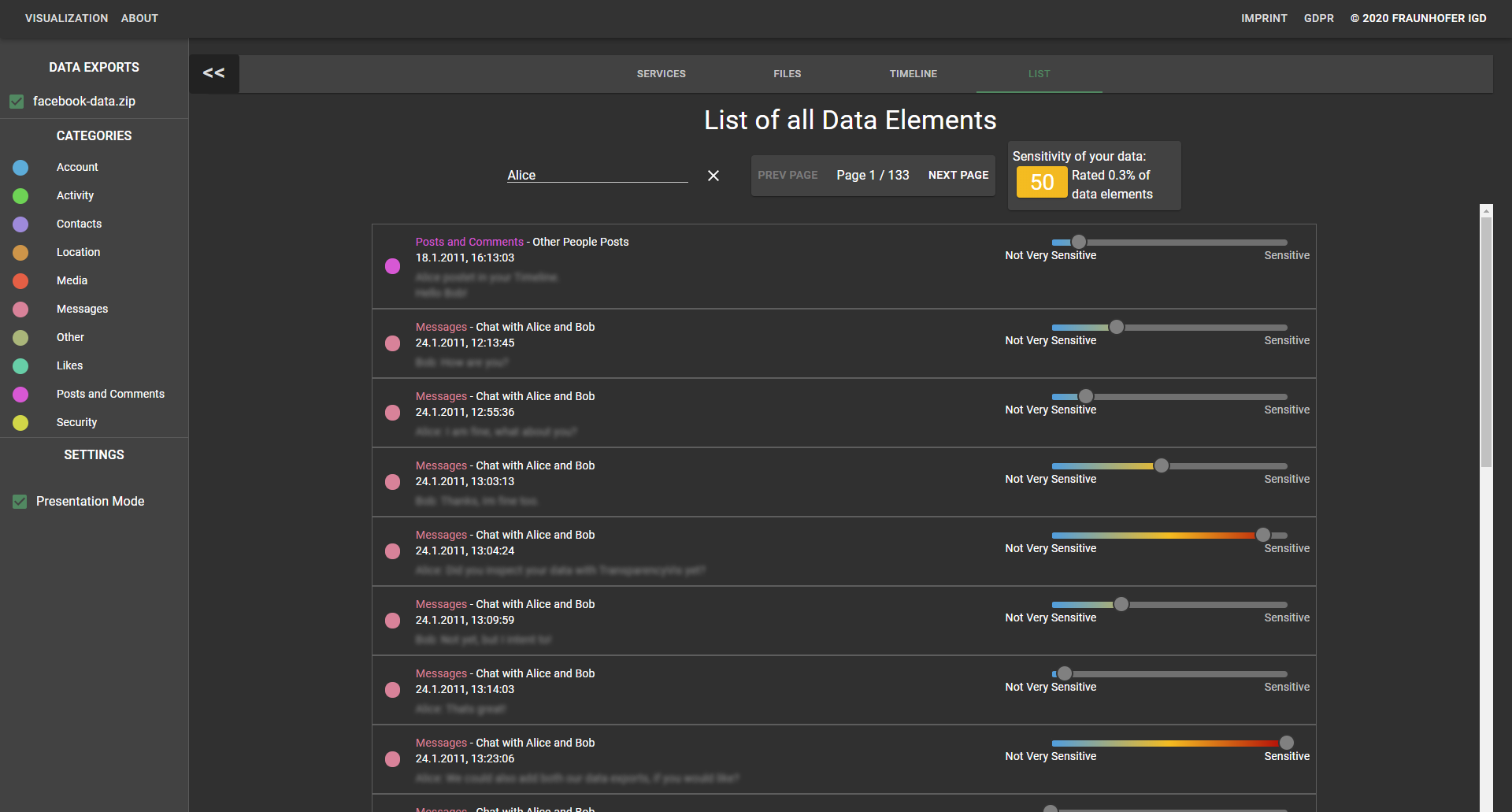}
        \caption{ListView}
     \label{fig:screenviews:list}
    \end{subfigure}
     \caption{The four views of \textit{TransparencyVis}, (a) \textit{Data Page} where the user can drag and drop his data export folder to, (b) \textit{FileView} gives an overview over all files contained in the export, (c) \textit{TimeView} with categorized data elements to explore temporal trends and patterns, (d) \textit{ListView} for details on each data element and the possibility to reflect on each data element by rating the perceived sensitivity.}
     \label{fig:screenviews}
\end{figure*}

% R3 and (Multiple Datasets) R6
At first, the user is provided with an initial view that consists of a dropzone to enter the data export and an overview of the supported services. 
The \textit{Data Page} (\autoref{fig:screenviews:data}) has a minimalistic design to reduce the users cognitive load.
For each service, an instruction on how to retrieve the data exports from the services is provided.
%The explanations can be switched between german and english.
% \brown{TODO: R3.2 Vielleicht sowas wie: demo dataset for Fundamentalists that might have no datasets available because of their distrust to data collection.}
% \red{A demo dataset is provided to test the application when no personal data export is available.}
After the data export is loaded to \textit{TransparencyVis}, the corresponding service field is colored and the inserted data set is listed within this field.
%\red{Multiple data exports can be loaded simultaneously.
After the data exports are processed they are kept in memory until the browser tab is closed or reloaded.

\subsubsection{FileView} \label{subsec:FileView}
The \textit{FileView} (\autoref{fig:screenviews:file}) displays all files of the data export in a treemap (\textbf{R1}).
We decided to use a treemap as our goal was to show an overview over all elements contained in the data export and to depict the proportions in the parts-to-a-whole relationship between the file elements and the whole export.
%TODO? As the folder structure exhibits a hierachical structure, a treemap was more appropriate to alternatives visualizations
Also we saw the metaphor of boxes, where the data elements are stored, as an appropriate representation for the file visualization.
Because the amounts of files contained in the export can vary much from user to user we see the space-filling treemap also as a good choice to support the scalability.
When choosing the treemap we also had the hierarchical data in mind.
While the current version only displays the leaves, for future extensions we aim to emphasize the hierarchical structure of the data to increase the understandability.
Each field represents a file which is contained in the export.
There are multiple attributes for the scaling of the treemap slices which the user can choose from.
As the main valuable options we see the file size and the amount of data points included in the files.
While the first attribute can help to discover large (possibly) sensitive files, as for example videos or images, the latter can help to discover collections of many elements.
This could for example be a conversation record or a search history with many items.
Details can further be inspected in the \textit{TimeView} or the \textit{ListView}.
Further possible but not yet implemented options would be to scale according to the sensitivity value.
However this option depends on the input from the user.
The color represents the category of the contained data (\textbf{R5}). 
Files which do not contain further data elements are colored white.
% Files, which contains datapoints, which can be visualized in the \textit{TimeView} have different colors,
% while data with no time attribute are white.
In the treemap a user can compare the different categories to see which is prevalent and how much data is collected in each category.
Users can inspect details about the files via tooltips and zooming (\textbf{R2}).
%\textcolor{red}{Users can select a file to zoom in and see details} (\textbf{R2}).
%Users can also change the mapping of the treemap to the amount of data elements or other attributes.
%This can, for example, be helpful to detect chats with a high amount of data elements, which reveals persons with whom one had many conversations.
Multiple data sets are merged in this view to one. 
This allows the user to combine the data from different services in one overview.
However, in the sidebar the user can select and deselect the data sets to display.
%\textcolor{brown}{TODO: FileView irgendwo groß als Bild}

\subsubsection{TimeView}
To support the exploration of patterns and trends, in this view, a timeline visualization (\autoref{fig:screenviews:time}) is used to display the temporal aspect of the data (\textbf{R4}).
Therefore, a scatterplot was chosen.
While time is one dimensional, the repetitive cycles are considered and split into two dimensions.
The x-axis shows the years and months across the data contained in the export.
The y-axis shows a single day.
%The x-axis represents the time axis across the years contained in the data export.
%The y-axis represents the time during the day, from 00:00 to 24:00.
A grid allows for better orientation and comparability.
%The time is defined in UTC and not converted to the local timezone as the users location is not always available in the given data.
Each circle in the visualization represents a data element.
To reduce overplotting, only a border of the circle is drawn.
The color indicates the category of the data elements.
We decided to use a scatterplot to allow a display of each single data element, while being able to perceive general trends.
Representing the data elements as units should support the perception of the possible relevance of every single data element. 
By assigning the data elements to a category and coloring them appropriately, the dense formation of the individual elements in the scatterplot additionally allows to observe patterns in groups of data elements.
To fulfill \textbf{R2} according to Shneidermans Mantra~\cite{shneiderman2003eyes} and support \textbf{R10} the familiar interaction paradigm \textit{zoom and pan} with the scroll wheel is implemented.
Therewith users can look at specific time frames, like years, months, or weeks by zooming into the timeline.
By seeing changes or deviations in the activity patterns it is possible for the users to identify certain important events in their life.
The data elements can be filtered based on the categories.
Therefore the user can click on the category filters on the left to hide irrelevant or overplotting data, such as the location history data from the Google service that is collected every few minutes on Android phones (\autoref{fig:teaser}, (4)).
To inspect the details of each data element, users can hover over the circles to view a tooltip that shows additional information about that item.
One extension of the current version after the evaluation was the search filed in \textit{TimeView} and \textit{ListView}.
With it, users can search for specific terms or names and inspect the patterns in a specific context.
Multiple datasets are merged by default and are shown in one timeline visualization. 
However, the \textit{MultiView} option allows the user to plot the different datasets on separate time lines. 
This is similar to small multiples and can be used to compare the patterns between different data exports.
This is shown in \autoref{fig:teaser} and is demonstrated in use case 2 in \autoref{sec:usecase2}.
A combination of multiple sources increases the possibilities to detect patterns such as daily routines, deviations, sleep, holidays, moving to a new place and others.

\subsubsection{ListView}
%\textcolor{blue}{Mehr ausführung zu Sensitivity, Berechnung des Scores diskutieren... }\todo{Steven/Jörn??}
The \textit{ListView} (\autoref{fig:screenviews:list}) is meant to support the user in inspecting the data elements in detail (\textbf{R2}) and in reflecting on the perceived sensitivity of this data, as required by \textbf{R7}.
It consists of a chronologically sorted scrollable list of the data elements from the selected data exports.
It displays the date, category and the contained text of each data element.
Further, the user has the possibility to rate the perceived sensitivity of each reviewed data point by interacting with a slider.
The slider allows to choose a value between \textit{Not very sensitive} and \textit{Very Sensitive}.
The average of the sensitivity rating over all elements is calculated and displayed to the user. 
This way the motivation to inspect and reflect on further data elements should be increased.
A search field can be used by the users to search for specific terms and thereby to inspect particular questions in detail.
The last two features are improvements based on the the evaluation results.

%\newpage
\section{Use Cases}\label{sec:usecases}

In this section, we demonstrate two use cases that show how \textit{TransparencyVis} can be used.
%the intendeted tasks could be performed by our prototype.
We do this by imaginary scenarios based on real data. 

\subsection{Use Case 1}  \label{sec:usecase1}
%- Overview, Details and Patterns}
%FieView + Overview + irgendwas... Detaillist (Search for something - for the Name of Messages) 
Bob has uploaded his data from \textit{Facebook} to \textit{TransparencyVis} by dragging and dropping the received zip archive into the \textit{Data Page} (\autoref{fig:screenviews:data}, \textbf{T3}).
In the \textit{FileView} (\autoref{fig:screenviews:file}) he can see all the files and folders contained in the data export (\textbf{T1}).
While hovering over the boxes and revealing the names of the files (\textbf{T2}), he wonders about some files, which he has sent to friends years ago and which seem to be still stored on \textit{Facebook's} servers (\textbf{T7}). 
%He also wonders about the actual amount of images, which are stored there - and which seems to be out of his control.
He also wonders about the large amount of images stored there, which he did not expect (or forgot about).
Then he spots the - in comparison to the others - relatively large message file (the big rose one). 
By inspecting the details in the tooltip, he learns that the file contains the conversation with Alice.
Having detected this, Bob might goes on to the \textit{TimeView} (\autoref{fig:screenviews:time}) and search for all data elements, which contain ``Alice''. In the timeline he can, for example, see that the conversation has mostly taken place around 2011 (\textbf{T6}).
But he also can explore further patterns of the conversation. 
Bob might go also to the \textit{ListView} (\autoref{fig:screenviews:list}) and search for ``Alice'' in the search field.
There he would get all messages which he has exchanged with her and could inspect, whether there is especially sensitive information, which he probably would like to delete.

 \begin{figure*} [htb]
         \includegraphics[clip, trim=0.9cm 6cm 0.9cm 4cm, width=1.00\textwidth]{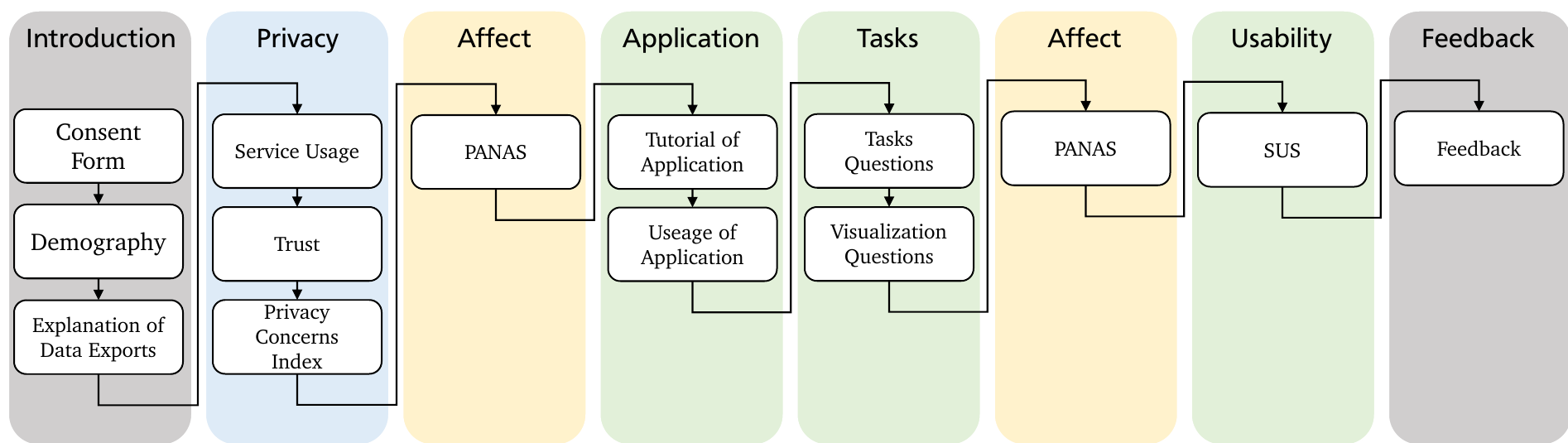}

          \caption{Evaluation proceeding - Questionnaire to receive feedback about \textit{evaluated user group}, \textit{appropriateness of TransparencyVis}, \textit{effect on the privacy  attitude} and \textit{usability}. The online evaluation started with an introduction part and followed by a questionnaire to derive users attitude to privacy. A PANAS questionnaire has been used to examine the participants emotional affect of seeing the data in \textit{TransparencyVis}. The participants could explore \textit{TransparencyVis} with their own data. The evaluation conclude with questions regarding usability and general feedback.}
          \label{fig:eval process}
  \end{figure*}
  
\subsection{Use Case 2}  \label{sec:usecase2}
%- Compare between datasets}
\autoref{fig:teaser} shows how four different datasets can be compared (\textbf{T5}) with each other in one view.
Alice (left) and Bob (right) have both provided their personal data sets retrieved from \textit{Facebook} (top) and from \textit{Google} (bottom) - (\textbf{T3}).
Compared to Bob, Alice seems to have used \textit{Facebook} mostly for private messaging (rose circles in (1)).
Hovering over the circles reveals the communication partner as well as the full message text of the message item (\textbf{T2}).
According to the data, Alice primarily used \textit{Facebook} (1) rather than \textit{Google} (3).
She seems to have some messaging data on \textit{Google} around 2015, but then she seems to have avoided using her \textit{Google} account (\textbf{T6}).
In contrast to this, Bob's \textit{Google} dataset (4) reveals a large amount of tracked activities (green). 
Beginning in 2014, his location is tracked constantly (orange).
Each circle reveals the concrete stored information in a tooltip, like actual location coordinates, search terms, seen videos or visited webpages.
Bob has an \textit{Android} phone, which is connected to his \textit{Google} account, while Bob's privacy settings allow \textit{Google} to track all of his activities on the platform. 
% \textcolor{brown}{to track everything they want./nearly all of his activities/... to track activities when using the service like search, video history, web history, location history and others}
Alice on the other hand was surprised to discover that the green activities around 2015 hint at her \textit{Youtube} history of videos she had watched at that time.
% \brown{She realized that she would feeld uncomftable to share this with others (\textbf{T7}).}
Thinking about how her taste and interests have changed over the years, she caught herself at the thought, that she would feel uncomfortable to share part of the history with others (\textbf{T7}).
Both Bob and Alice noticed that security related data in the collections of \textit{Facebook} has increased since around 2016.
While the scenario that two users would provide their data to merge them in one visualization, seems quite unrealistic, we decided for this use case to present the possibilities of the tool and to present the difference of data sets between different personalities.
However possible applications of this scenario might be, the combination of data sets of members of a family or the comparison of own data with exemplary average datasets.

%%%%%%%%%%%%%%%%%%%%%%%%%%%%%%%%%%%%%%%%%%%%%%%%%%%%%%%%%%%%%%%%
\section{Evaluation}
%%%%%%%%%%%%%%%%%%%%%%%%%%%%%%%%%%%%%%%%%%%%%%%%%%%%%%%%%%%%%%%%

\begin{figure*}[htb!]
    \centering
    \begin{subfigure}[b]{0.32\textwidth}
        \includegraphics[clip, trim=1.7cm 8.9cm 1.7cm 10.4cm, width=1.00\textwidth]{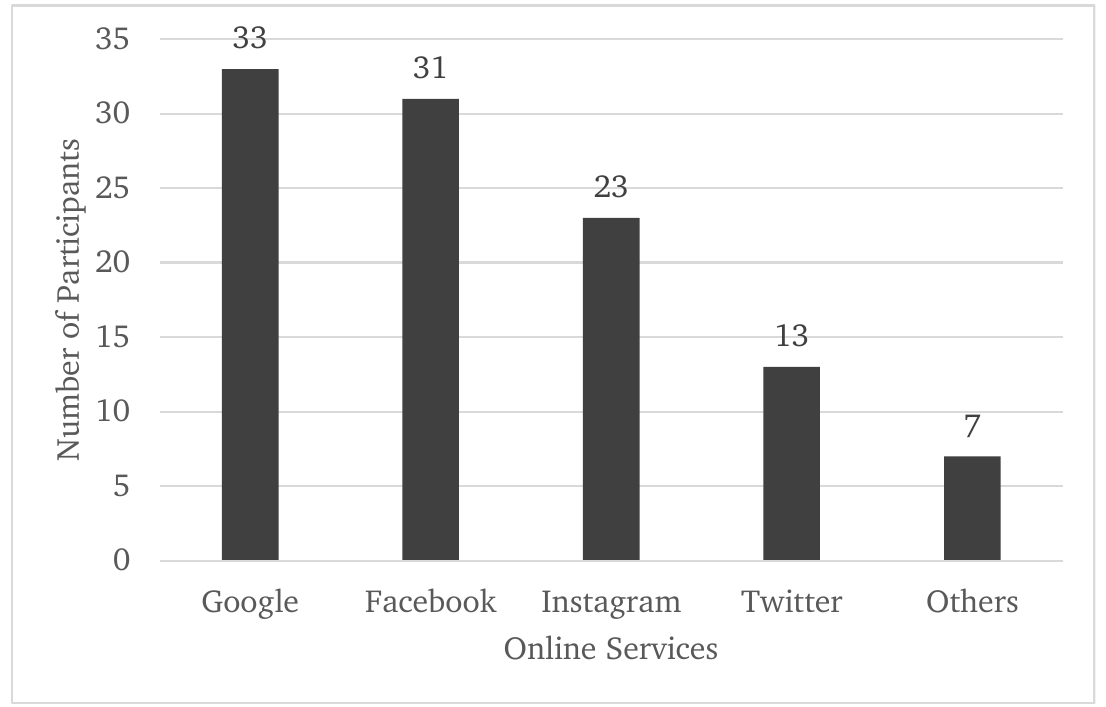}
        \caption{used online services}
        \label{fig:eval results:a}
    \end{subfigure}
    \begin{subfigure}[b]{0.32\textwidth}
        \includegraphics[clip, trim=1.7cm 8.9cm 1.7cm 10.4cm, width=1.00\textwidth]{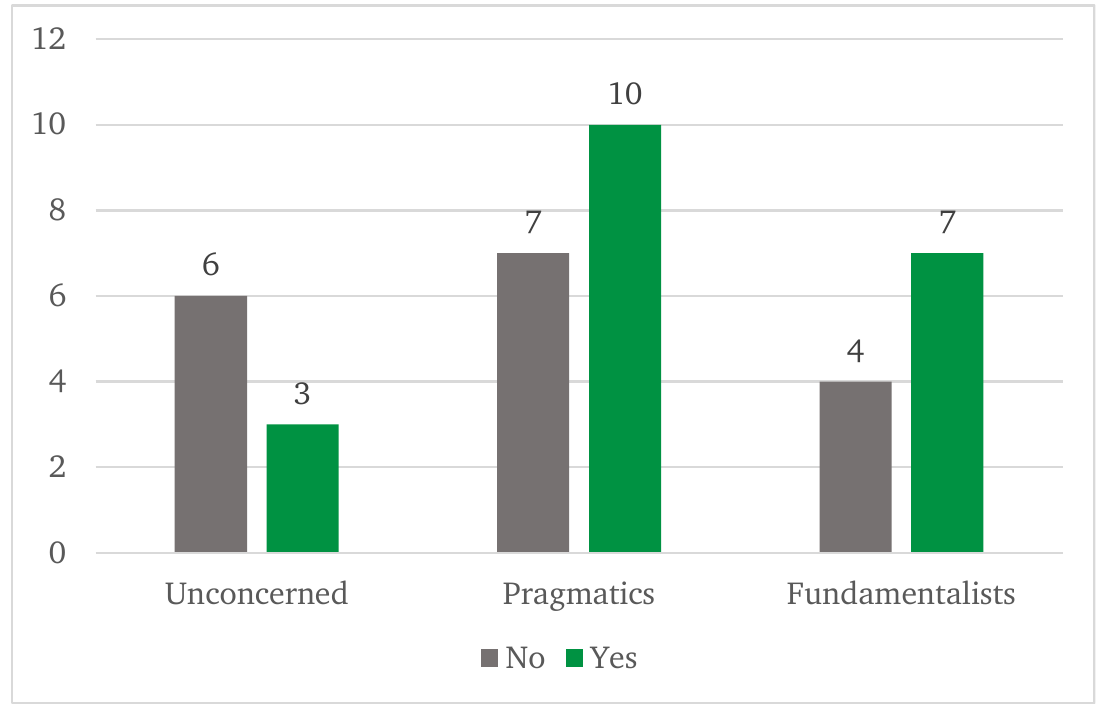}
        \caption{saw trends}
        \label{fig:eval results:b}
    \end{subfigure}
    \begin{subfigure}[b]{0.32\textwidth}
        \includegraphics[clip, trim=1.7cm 8.9cm 1.7cm 10.4cm, width=1.00\textwidth]{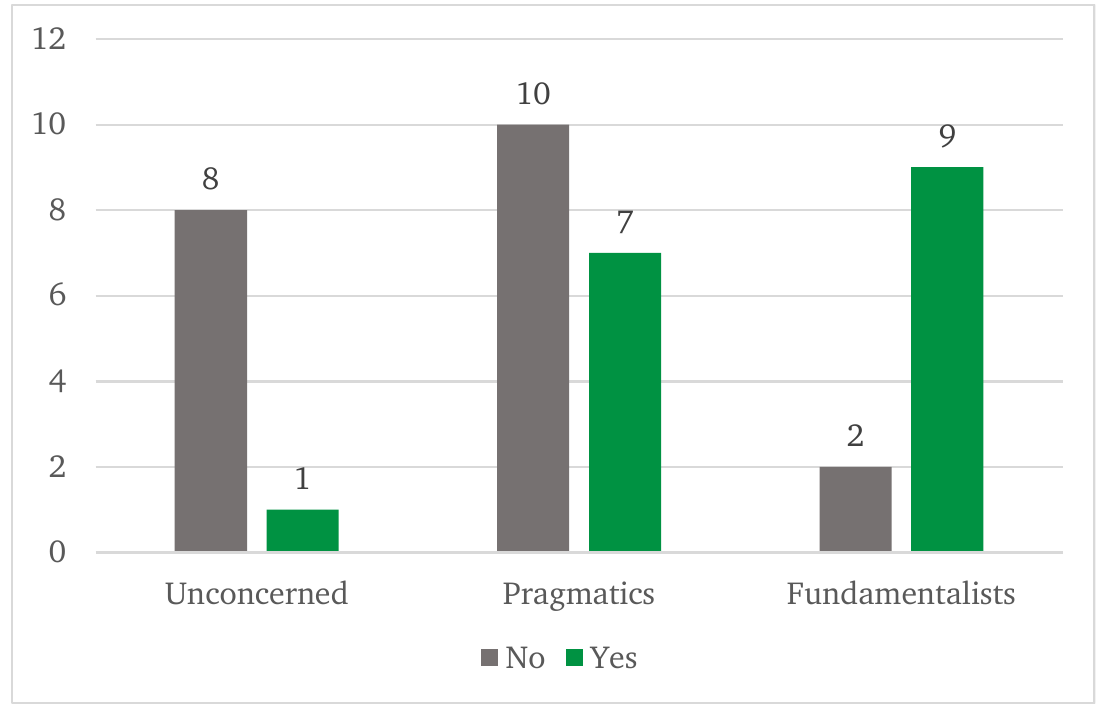}
        \caption{changed privacy attitude}
        \label{fig:eval results:c}
    \end{subfigure}
        \caption{Evaluation results. (a) Amount of participants using each online service, (b) Amount of participants that claimed to have seen any trends in their data, differentiated according to the privacy attitude groups, (c) Amount of participants, who claimed to have changed their privacy attitude after having seen their data through \textit{TransparencyVis} }
     \label{fig:eval results}
\end{figure*}

We evaluated the first iteration of our prototype \textit{TransparencyVis}, with regard to the three goals defined in \autoref{sec:tasks}. The evaluation focused on four main aspects: \textit{(1) evaluated user group}, \textit{(2) appropriateness of TransparencyVis}, \textit{(3) effect on the privacy attitude} and \textit{(4) usability}.
We have used the results to improve \textit{TransparencyVis} into the version presented in this paper.

\subsection{Methodology}
As detailed in \autoref{sec:transparencyvis}, the prototype is a web interface that can be used with personal data.
Therefore, we have conducted an online study with 37 users (14 f, 21 m, 2 other) and their own personal data. 
The age ranged between 20 and 64 years with a predominance on the age group of 20-34 years (30/37).
Most participants were either students (12) or employees (24). 
All participants were from Germany.
% ... später However, we also provided a demo data set.
The study ran for 21 days.
The average duration of an evaluation session was about 30 minutes.
Only the participants that reached the last page of the evaluation were recorded.
The participants were led through a fixed process by the evaluation tool~\cite{soscisurvey} without the need for an instructor. 
Therefore, participants could conduct the evaluation on their own, in their own pace and in their familiar environment.
This way the usage of the tool during the evaluation leaned on the natural context of an every day situation, in line with the targeted user group.
The process of the evaluation is shown in \autoref{fig:eval process}.
The questions of the evaluation can be found in supplementary materials.
%ÜBERLEGEN, ob man das einfach in die CAPTION tut und sich diesen ganzen Absatz spart.
The evaluation started with the introduction, which consists of a consent form, a questionnaire on demographic data and a data preparation session. 
Then, the participants had to fill out a questionnaire about their attitude towards privacy. 
This questionnaire was inspired by the works of Cabinakova et al.~\cite{cabinakova2016empirical} (trust), Westin~\cite{westin1991harris} and Bergmann~\cite{bergmann2008testing} (privacy concern index).
Furthermore, we asked the participants to fill out the PANAS questionnaire~\cite{watson1988development} before and after the actual interaction with the tool.
This was used to measure the possible emotional affect caused by the exploration of the own data as provided by \textit{TransparencyVis}.
After using the tool, some questions regarding possible discoveries were asked, followed by a questionnaire about the perceived appropriateness of \textit{TransprencyVis} for some selected tasks, primarily concerning the goals to support \textit{perception} and \textit{comprehension}.
Then we checked the overall perceived usability with the SUS questionnaire~\cite{bangor2009determining}.
Finally, we asked the participants for their subjective opinion, if and how the insights in the data have changed their attitude towards privacy and gathered more general feedback.

%% RESULTS %%%%
\subsection{Results}
\subsubsection{Evaluated user group}
In the set of participants were 9 \textit{Unconcerend}, 17 \textit{Pragmatics} and 11 \textit{Fundamentalists}, which goes along with the distributions observed by Bergmann~\cite{bergmann2008testing}, that unconcerned users are usually underrepresented.
The users' trust in services was measured with two questions from Cabinakova et al.~\cite{cabinakova2016empirical}.
The answers were converted from their Likert scale to a score from 0 to 100.
The mean of all participants was 68.2 with a standard deviation of 26.5.
% % correlation
% The trust of the participants and the $px$ value show a strong correlation of 0.61 using the Pearson correlation.
% A scatterplot with a trendline of $px$ and trust value is shown in [IMG] %\ref{chart:eval:trustandpx}.
% % number of services
Most users had an account on the \textit{Google} platform with 33 out of 36 participants, \textit{Facebook} with 30, \textit{Instagram} with 22 and \textit{Twitter} with 12 participants (see \autoref{fig:eval results}).
% providers per participant
%Participants were asked how many services they use and for how long.
% unconcerned
Participants from the \textit{Unconcerned} group used the most services with 2 to 5 services.
The \textit{Pragmatist} group used between 2 to 4 services.
The \textit{Fundamentalists} group used the least with 1 to 3 services.
% duration
The \textit{Unconcerned} had the least amount of hours used with an average of 15 hours weekly across all services.
The \textit{Pragmatists} group had an average of 32 hours, and the \textit{Fundamentalists} an average of 16 hours.
%
%outro to next chapter
In conclusion, the participants of this survey are well distributed in their privacy attitude and users of multiple services.

\subsubsection{Appropriateness of the tool}
Overall, we have received much appreciation by the participants as well as from informal presentations of the interface.
Nine participants expressed their praise explicitly in the feedback section with an appropriate comment.
Several participants asked if they could forward the link to friends.\\
% \paragraph{G1: Support perception of data}
%\textbf{G1: Support perception of data:} 
\textbf{Support perception of data:} 
The questionnaire supports the appropriateness of the tool for the perception of the \textit{amount of data} (28/37 agreed), the \textit{type of data} (25/37 agreed) and \textit{trends and patterns} (18/37 agreed). 
These are the main aspects with regard to the goal to support \textit{perception}\\
%\textbf{G1} (\textit{Perception}).\\
%Our funtion to rate the perceived sensitivity of selected data items has not been approved ...
%\textcolor{red}{ggf noch was zur feedback function sagen...}
%\paragraph{G2: Support comprehension of data}
%\textbf{G2: Support comprehension of data:} 
\textbf{Support comprehension of data:} 
To determine whether the participants were able to bring the perceived data in context with their meaning, we asked what they saw during the exploration phase.
This way we wanted to estimate the effect with respect to the goal to support \textit{comprehension}.
%\textbf{G2} (\textit{Comprehension}).
In particular we asked about the patterns or insights that participants have gained from their data.
Interestingly, the ability to find patterns seems to relate with the privacy concern group to which the participant belonged.
This is shown in \autoref{fig:eval process}.
While the majority of the \textit{Fundamentalists} (7/11) reported about exciting trends, only 3 of 9 \textit{Unconcerned} have claimed to see any trend. 
For the \textit{Pragmatists} it appeared to be half-half.
We received reports about some identified trends, which we have clustered in the following groups.

\begin{enumerate} \itemsep-0.5ex
    \item \textbf{Usage patterns regarding the platform:} e.g decreased activity on \textit{Facebook}, changes from one platform to another
    \item \textbf{Changes in online behavior:} Periods of a high amount of messages or friends requests, last deletion of the browser-history and similar
    \item \textbf{Changes in location:} Changes of place of residency, change of workplace, holidays, location change from working day to weekend
    \item \textbf{Changes induced by the platform:} For example increase of security elements from \textit{Facebook} in recent years
    \item \textbf{Personal events and patterns:} Sleeping patterns, online times, holidays, birthday congratulations, or change of jobs
\end{enumerate} 

\subsubsection{Usability and improvements}
The SUS revealed an average score of 65.4, which is a good value for the first iteration. 
There was nearly no difference between the different user groups.
We further clustered the textual answers from the feedback section and derived the following main points for improvement,
which we have adjusted in the version presented in this paper.
\begin{enumerate} \itemsep-0.5ex
        \item \textbf{Zoom function:} Adding a zoom function or a selection of a time-period to the TimeView  
        \item \textbf{Filter improvement:} Improve the filter option, e.g. filtering on the categories, reducing the overload 
        \item \textbf{Tooltip improvement:} e.g. format, details, position  
        \item \textbf{Search function:} Improving the support for pattern detection by adding a search capability to the timeline
\end{enumerate}

Furthermore, \textit{FileView} and the \textit{ListView} turned out to be complicated to understand for the participants.
We have implemented some improvements for the current version.
For the \textit{FileView} we have simplified some interactions and improved the tooltips.
We further classified the files (white) additionally according to the type of file, which is displayed as a label and in the tooltip.
We also have simplified the layout of the \textit{ListView} and added the average rating value as a feedback for rating of the perceived sensitivity. Additionally we added a search functionality to filter the elements.

Summarized, the main extensions we have implemented after the evaluations are: Support of multiple data sets simultaneously, filter by categories, zoom and pan, feedback for the sensitivity rating, improved tooltips and layout simplifications.

\subsubsection{Effect on privacy attitude} 
With regard to the goal to support \textit{projection}, we wanted to know whether the use of \textit{TransparencyVis} had any effect on the privacy attitude of the participants. 
The results of the PANAS questionnaire revealed a significant increase of the negative attributes \textit{Upset} (+0,78) and \textit{Scared} (+0,65) and a marginal significant loss of the positive attribute \textit{Determined} (-0.41).
%According to this the participants seemed to be more scared and upset and also less determined after seeing their data trough \textit{TransparencyVis}. 
With one of our goals being to trigger more attention for the 
%own
effects of online behavior, an increase of a slight alertness based on the insights can be seen as success.
However, while this evaluation only meant to get a trend about possible effects, further studies should conduct deeper evaluations on the effects and their reasons.\\
\textbf{Support projection:} 
17 participants confirmed that the use of \textit{TransparencyVis} had an influence on their privacy attitude.
Interestingly, most of these participants belonged to the group \textit{Fundamentalists} (9/11), while only one \textit{Unconcerned} (1/9) was affected in a similar way (see \autoref{fig:eval results:b}).
Further we clustered the answers to the questions about which kind of influence has been experienced.
%Further we clustered the answers to the questions wether the use of TransaprencyVis had an influence on the privacy attitude of the proband and if yes, which kind of influence.
Overall, we derived the following clusters of answers to the questions on privacy attitude:

\begin{enumerate} \itemsep-0.5ex
    \item \textbf{More attentiveness:}  Many anticipated on more attentiveness for their own personal data handling and online behavior (8 participants)
    \item \textbf{Checking settings:} Some intended to check and change privacy settings, maybe switch to more trustful platforms (4)
    \item \textbf{Deletion:} Some stated to delete their data, the entire account or avoiding such platforms (4)
    \item \textbf{Gain of Confidence:} Selected participants increased their confidence in treating their personal online data (2)
    \item \textbf{Surprise:} Some expressed surprise about which data actually has been collected (2)
    \item \textbf{Curiosity:} One expressed curiosity about what the exported copy might not include (1)
\end{enumerate}
%%%%%%%%%%%%%%%%%%%%%%%%%%%%%%%%%%%%%%%%%%%%%%%%%%%%%%%%%%%%%%%%
%\section{Redesigned Version (optional)}
%%%%%%%%%%%%%%%%%%%%%%%%%%%%%%%%%%%%%%%%%%%%%%%%%%%%%%%%%%%%%%%%

%%%%%%%%%%%%%%%%%%%%%%%%%%%%%%%%%%%%%%%%%%%%%%%%%%%%%%%%%%%%%%%%
\section{Discussion}
%%%%%%%%%%%%%%%%%%%%%%%%%%%%%%%%%%%%%%%%%%%%%%%%%%%%%%%%%%%%%%%%
With our design study, we have gained several insights into the understanding of personal data stored by online services.
First, the idea and the current implementation have received much approval and interest from the targeted user group.
We have observed, that especially the inclusion of the usability requirements \textbf{R8-R10} had a strong influence on the positive feedback.
This is especially true, because the users could use the tool with their own personal data at their own pace in their own private environment. 
The evaluation showed good results when reflecting on the effect on the privacy attitude and perceived appropriateness of the tool for the intended purposes.
However, the obviously subjective answers with regard to the change in privacy attitude should not be seen as possible trends in changed behavior.
Therefore, long-term studies have to be applied on improved versions of the interface to examine the significance of the effects.
During the evaluation we have also gained valuable feedback, on how to improve the usability of the interface, which we have partly integrated into the version of the tool presented in this paper.
%Some implications have turned out to require further research.
Some implications will require further research.
% \brown{This is especially true for the \textit{FileView} and the \textit{ListView}.(Fillersatz, einfach direkt die beiden im nächsten nennen)}
%, which (the later) originally was ment to encourage the user to reflect on the perceived sensitivity of the data points.
This is especially true for \textit{ListView} and \textit{FileView}, which purpose seemed to have not been understood very well by the participants.
% \textcolor{red}{While we have already done some optimizations to the \textit{ListView} as presented in this paper}, 
Additionally to the optimizations in this paper, a stronger improvement of the rating functionality and the appropriate feedback of the \textit{ListView} should be carried out in future work. 
Especially the calculation of the perceived sensitivity should get a stronger attention.
While the concept of the treemap in the \textit{FileView} seems to be not very intuitive for the common users, it has many advantages with regard to the data of the exports, as described in \autoref{subsec:FileView}. Future work, however, should take the optimization of the treemap visualization for non-experts into account.
% \green{TODO: treemap is bad because blabla, but we wanted blabla, and blabla is more important. because of that we choose treemap in the end.} 
%\green{TODO: Kurze DIskussion über Sensitivity und ListView} 
One of the challenges, which is also related to the \textit{FileView}, is the huge variance in the formats of the data exports between different online services as well as between different users.
This complicates the development of appropriate parsers for the proposed unification scheme.
%\red{This makes the parsing and the categorization of the files difficult. (??)}
%Furthermore, the files often don't have very meaningful names, leads to a large amount of  cryptic names and undefined format.
%In the current version of the prototype, this results in a large amount of files that are classified as ``Other format'' and have cryptic names, which is not \red{really valuable\todo{anderes wort?}} for the user.
The problem of the huge variance in formats and content also leads to the open challenge, to achieve a comprehensive overview for the user.
Additionally the communication of the difference between \textit{files} and \textit{data elements} to the user still needs to be improved.

Overall, we are encouraged by the results of the evaluation of the first iteration of the tool and are more confident that tools of this type have the potential to receive attention by a broad range of Internet users. 
While the current version is primarily designed as an independent interface for the individual, the application of such a visualization by online services is another possibility.
Such functionality could increase the users' trust in the services, which is an increasingly important factor for the willingness to share personal data with an online service ~\cite{norberg2007privacy, cabinakova2016empirical}.
%Trust is the main indicator in a users' desire and behavior to share personal data with a service~\cite{norberg2007privacy, cabinakova2016empirical}.
As an extended stand-alone application, the interface could, however, also be used as a management tool for online data by bringing the exports of all used online services together.

%%%%%%%%%%%%%%%%%%%%%%%%%%%%%%%%%%%%%%%%%%%%%%%%%%%%%%%%%%%%%%%%
\section{Conclusion and Future Work}
%%%%%%%%%%%%%%%%%%%%%%%%%%%%%%%%%%%%%%%%%%%%%%%%%%%%%%%%%%%%%%%%
In this paper, we have presented our design study on increasing the attention and awareness of the common internet user for their own personal data that are stored by different online services.
We have presented the targeted user group, which we differentiated by the privacy concern index together with the used data source.
We also have provided a unification scheme based on two defined data types and ten plus one categories, which can be used by other researchers to develop new parsers for further services.
We have also presented the tasks which we have derived based on the theory of situation awareness applied to stored personal data.
Based on the derived requirements we have implemented the online accessible prototype \textit{TransparencyVis}, which can be used with own real personal data.
We have evaluated this tool with 37 targeted users and have elicited important insights with regard to the tool's appropriateness, usability and effect on the participant's attitude towards privacy.
%%%%%%%%%%%%%%%%%%%%%%%%%%%%%%%%%%%%%%%%%%%%%%%%%%%%%%%%%%%%%%%%
%\section{Future Work}
%%%%%%%%%%%%%%%%%%%%%%%%%%%%%%%%%%%%%%%%%%%%%%%%%%%%%%%%%%%%%%%%
The evaluation of this first iteration has led to many ideas for improvements of the approach.
The main next steps would be to improve the \textit{FileView} to enhance the overview of all data contained in the download at a first glance. 
Further we want to investigate how an active reflection on the presented data can be supported more effectively.
A possible approach could be to connect a users rating of the sensitivity to an active learning approach to support the visualization of the results.
A remaining challenge for any new ideas is our effort to preserve the privacy of the user by not using a server-based approach.
Further potential improvements include the employment of additional analysis methods, including other services, and further optimizations of the data parsing.

\acknowledgments{
This research work has been funded by the German Federal Ministry of Education and Research and the Hessen State Ministry for Higher Education, Research and the Arts within their joint support of the National Research Center for Applied Cybersecurity ATHENE.}

\bibliographystyle{abbrv-doi}
%\bibliographystyle{abbrv-doi-narrow}
%\bibliographystyle{abbrv-doi-hyperref}
%\bibliographystyle{abbrv-doi-hyperref-narrow}

%\bibliography{template}
\bibliography{Bibliography}

% \section{Appendix}

\end{document}